\documentstyle[twoside,psfig]{article}

\catcode`\@=11
\long\def\@makefntext#1{
\protect\noindent \hbox to 3.2pt {\hskip-.9pt
$^{{\eightrm\@thefnmark}}$\hfil}#1\hfill}  

\def\thefootnote{\fnsymbol{footnote}}
\def\@makefnmark{\hbox to 0pt{$^{\@thefnmark}$\hss}} 

\def\ps@myheadings{\let\@mkboth\@gobbletwo
\def\@oddhead{\hbox{}
\rightmark\hfil\eightrm\thepage}
\def\@oddfoot{}\def\@evenhead{\eightrm\thepage\hfil
\leftmark\hbox{}}\def\@evenfoot{}
\def\sectionmark##1{}\def\subsectionmark##1{}}

\oddsidemargin=\evensidemargin
\renewcommand{\thefootnote}{\fnsymbol{footnote}}

\newcounter{sectionc}\newcounter{subsectionc}\newcounter{subsubsectionc}
\renewcommand{\section}[1] {\vspace{12pt}\addtocounter{sectionc}{1}
\setcounter{subsectionc}{0}\setcounter{subsubsectionc}{0}\noindent
 {\tenbf\thesectionc. #1}\par\vspace{5pt}}
\renewcommand{\subsection}[1] {\vspace{12pt}\addtocounter{subsectionc}{1}
 \setcounter{subsubsectionc}{0}\noindent
 {\bf\thesectionc.\thesubsectionc. {\kern1pt \bfit #1}}\par\vspace{5pt}}
\renewcommand{\subsubsection}[1]
{\vspace{12pt}\addtocounter{subsubsectionc}{1}
 \noindent{\tenrm\thesectionc.\thesubsectionc.\thesubsubsectionc.
 {\kern1pt \tenit #1}}\par\vspace{5pt}}
\newcommand{\nonumsection}[1] {\vspace{12pt}\noindent{\tenbf #1}
 \par\vspace{5pt}}

\newcounter{appendixc}
\newcounter{subappendixc}[appendixc]
\newcounter{subsubappendixc}[subappendixc]
\renewcommand{\thesubappendixc}{\Alph{appendixc}.\arabic{subappendixc}}
\renewcommand{\thesubsubappendixc}
 {\Alph{appendixc}.\arabic{subappendixc}.\arabic{subsubappendixc}}

\renewcommand{\appendix}[1] {\vspace{12pt}
        \refstepcounter{appendixc}
        \setcounter{figure}{0}
        \setcounter{table}{0}
        \setcounter{lemma}{0}
        \setcounter{theorem}{0}
        \setcounter{corollary}{0}
        \setcounter{definition}{0}
        \setcounter{equation}{0}
        \renewcommand{\thefigure}{\Alph{appendixc}.\arabic{figure}}
        \renewcommand{\thetable}{\Alph{appendixc}.\arabic{table}}
        \renewcommand{\theappendixc}{\Alph{appendixc}}
        \renewcommand{\thelemma}{\Alph{appendixc}.\arabic{lemma}}
        \renewcommand{\thetheorem}{\Alph{appendixc}.\arabic{theorem}}
        \renewcommand{\thedefinition}{\Alph{appendixc}.\arabic{definition}}
        \renewcommand{\thecorollary}{\Alph{appendixc}.\arabic{corollary}}
        \renewcommand{\theequation}{\Alph{appendixc}.\arabic{equation}}
        \noindent{\tenbf Appendix \theappendixc #1}\par\vspace{5pt}}
\newcommand{\subappendix}[1] {\vspace{12pt}
        \refstepcounter{subappendixc}
        \noindent{\bf Appendix \thesubappendixc. {\kern1pt \bfit #1}}
 \par\vspace{5pt}}
\newcommand{\subsubappendix}[1] {\vspace{12pt}
        \refstepcounter{subsubappendixc}
        \noindent{\rm Appendix \thesubsubappendixc. {\kern1pt \tenit #1}}
 \par\vspace{5pt}}

\newcommand{\textlineskip}{\baselineskip=13pt}
\newcommand{\smalllineskip}{\baselineskip=10pt}

\def\eightcirc{
\begin{picture}(0,0)
\put(4.4,1.8){\circle{6.5}}
\end{picture}}
\def\eightcopyright{\eightcirc\kern2.7pt\hbox{\eightrm c}}

\newcommand{\copyrightheading}[1]
 {\vspace*{-2.5cm}\smalllineskip{\flushleft
 {\footnotesize International Journal of Modern Physics A #1}\\
 {\footnotesize $\eightcopyright$\, World Scientific Publishing
  Company}\\
  }}

\newcommand{\publisher}[2]{{\begin{center}\footnotesize\smalllineskip
 Received #1\\
 Revised #2
 \end{center}
 }}

\def\abstracts#1#2#3{{
 \centering{\begin{minipage}{4.5in}\footnotesize\baselineskip=10pt
 \parindent=0pt #1\par
 \parindent=15pt #2\par
 \parindent=15pt #3
 \end{minipage}}\par}}

\newcommand{\bibit}{\nineit}

\renewenvironment{thebibliography}[1]
 {\frenchspacing
  \ninerm\baselineskip=11pt
  \begin{list}{\arabic{enumi}.}
 {\usecounter{enumi}\setlength{\parsep}{0pt}
  \setlength{\leftmargin 12.7pt}{\rightmargin 0pt} 
  \setlength{\itemsep}{0pt} \settowidth
 {\labelwidth}{#1.}\sloppy}}{\end{list}}

\newcounter{itemlistc}
\newcounter{romanlistc}
\newcounter{alphlistc}
\newcounter{arabiclistc}
\newenvironment{itemlist}
     {\setcounter{itemlistc}{0}
  \begin{list}{$\bullet$}
 {\usecounter{itemlistc}
  \setlength{\parsep}{0pt}
  \setlength{\itemsep}{0pt}}}{\end{list}}

\newcommand{\fcaption}[1]{
        \refstepcounter{figure}
        \setbox\@tempboxa = \hbox{\footnotesize Fig.~\thefigure. #1}
        \ifdim \wd\@tempboxa > 5in
           {\begin{center}
        \parbox{5in}{\footnotesize\smalllineskip Fig.~\thefigure. #1}
            \end{center}}
        \else
             {\begin{center}
             {\footnotesize Fig.~\thefigure. #1}
              \end{center}}
        \fi}

\newcommand{\tcaption}[1]{
        \refstepcounter{table}
        \setbox\@tempboxa = \hbox{\footnotesize Table~\thetable. #1}
        \ifdim \wd\@tempboxa > 5in
           {\begin{center}
        \parbox{5in}{\footnotesize\smalllineskip Table~\thetable. #1}
            \end{center}}
        \else
             {\begin{center}
             {\footnotesize Table~\thetable. #1}
              \end{center}}
        \fi}

\def\@citex[#1]#2{\if@filesw\immediate\write\@auxout
 {\string\citation{#2}}\fi
\def\@citea{}\@cite{\@for\@citeb:=#2\do
 {\@citea\def\@citea{,}\@ifundefined
 {b@\@citeb}{{\bf ?}\@warning
 {Citation `\@citeb' on page \thepage \space undefined}}
 {\csname b@\@citeb\endcsname}}}{#1}}

\newif\if@cghi
\def\cite{\@cghitrue\@ifnextchar [{\@tempswatrue
 \@citex}{\@tempswafalse\@citex[]}}
\def\citelow{\@cghifalse\@ifnextchar [{\@tempswatrue
 \@citex}{\@tempswafalse\@citex[]}}
\def\@cite#1#2{{$\null^{#1}$\if@tempswa\typeout
 {IJCGA warning: optional citation argument
 ignored: `#2'} \fi}}

\def\pmb#1{\setbox0=\hbox{#1}
 \kern-.025em\copy0\kern-\wd0
 \kern.05em\copy0\kern-\wd0
 \kern-.025em\raise.0433em\box0}

\def\fnt#1#2{\footnotetext{\kern-.3em
 {$^{\mbox{\scriptsize #1}}$}{#2}}}

\def\thefootnote{\fnsymbol{footnote}}
\def\@makefnmark{\hbox to 0pt{$^{\@thefnmark}$\hss}} 

\def\ps@myheadings{%
    \let\@oddfoot\@empty\let\@evenfoot\@empty
    \def\@evenhead{\slshape\leftmark\hfil}
    \def\@oddhead{\hfil{\slshape\rightmark}}
    \let\@mkboth\@gobbletwo
    \let\sectionmark\@gobble
    \let\subsectionmark\@gobble
    }
\font\tenrm=cmr10
\font\tenit=cmti10
\font\tenbf=cmbx10
\font\bfit=cmbxti10 at 10pt
\font\ninerm=cmr9
\font\nineit=cmti9

\font\eightrm=cmr8

\def\qed{\hbox{${\vcenter{\vbox{   
   \hrule height 0.4pt\hbox{\vrule width 0.4pt height 6pt
   \kern5pt\vrule width 0.4pt}\hrule height 0.4pt}}}$}}

\renewcommand{\thefootnote}{\fnsymbol{footnote}}  

\begin{document}
\setlength{\textheight}{7.7truein}  

\thispagestyle{empty}

\markboth{\protect{\footnotesize\it Casimir Energy
and Vacua vor Superconducting Ball}}{\protect{\footnotesize\it Casimir
Energy and Vacua for Superconducting Ball}}

\normalsize\textlineskip

\setcounter{page}{1}

\copyrightheading{}  {Vol.~17, No.~6\&7 (2002) 920--925}

\vspace*{0.88truein}

\centerline{\bf CASIMIR ENERGY AND VACUA FOR }
\vspace*{0.035truein}
\centerline{\bf  SUPERCONDUCTING BALL IN SUPERGRAVITY}
\vspace*{0.37truein}
\centerline{\footnotesize ALEXANDER BURINSKII}
\centerline{\footnotesize\it Gravity Research
Group, NSI, Russian Academy of Sciences, B. Tulskaya 52.}
\baselineskip=12pt
\centerline{\footnotesize\it
 Moscow 113191, Russia}
\vspace*{10pt}
\vspace*{0.225truein}
\publisher{(received date)}{(revised date)}

\vspace*{0.21truein}
\abstracts{Casimir energy for solid conducting ball is considered on the
base of some finite models. One model is physical and built of a battery
of parallel metallic plates. Two finite models are based on the Higgs
model of superconductivity.
One of them is supersymmetric and  based on the Witten field model for
superconducting strings.
Treatment shows that contribution of Casimir energy can be very essential
for superdence state in the neutron stars and nuclear matter.}
{}{}


\vspace*{1pt}\textlineskip 
\section{Introduction} 
\vspace*{-0.5pt}
\noindent
\setcounter{footnote}{0}
\renewcommand{\thefootnote}{\alph{footnote}}
Starting from the first paper by Casimir \cite{Cas} on this subject,
the problem of vacuum energy for  (super)conducting ball
acquires the troubles caused by divergences by use the
perturbative approaches. The divergent vacuum volume term
$2\times \frac 12 \sum_{k} \omega_k$ is usually ignored in the field theory
and when determining the Casimir effect with plates and shells.
Meanwhile, for island systems there "appears some doubt" \cite{IZ,BD}
concerning the correctness of the removal of the zero-point term by
the way of traditional regularization procedures \cite{BD}.
Casimir was the first who pointed out on the possible effect of the
volume vacuum term. In \cite{Cas}, analyzing the classical Dirac electron
model, he considered two versions of the effect: {\it volume effect} with
conducting solid ball and {\it surface effect} with conducting sphere.
In the case of
sphere the vacuum zero-point field exists outside the sphere as
well as inside.
In the case of ideal conducting ball the interior vacuum modes must be
entirely excluded, resulting in the infinite
jump of the vacuum energy density on the surface of the ball
\begin{equation}
\Delta E_{vac} = E_{in} - E_{out} =
0 - V^{-1} \sum_{k}^{\infty} \omega_k =
- \lim_{k_{max}\to\infty} k_{max}^4 /8\pi ^2.  \label {E0vac}
\end{equation}

The necessity of the explicit introduction of the cut-off parameter
$k_{max}$
compelled Casimir to abandon this volume effect and prefer the effect with
sphere, where the above divergence is cancelled.

We consider here some related to this problem models where the result
turns out to be finite,  and we are going to conclusion that contribution of
the Casimir energy can be very essential for superdense matter in the
neutron
stars and nuclear systems.

\par
\section{Some models leading to finite Casimir energy for conducting ball}

\subsection{ Interatomic separation as a cut-off parameter }

A natural cut-off parameter is connected with
the real properties of material since the real conductors prove to be
transparent for high-frequency modes with $k>k_{max}\sim \pi/d$, where $d$
is
of order of interatomic separation. Consequently, one determines the jump
(\ref{E0vac}) of the vacuum field energy, the low-frequency terms must
be taken into account only. It leads to finite result
$\Delta E_{vac} = E_{in} - E_{out} =
V^{-1} \sum_{k}^{k_{max}} \omega_k
 - V^{-1} \sum_{k}^{\infty} \omega_k =
 - V^{-1} \sum_{k}^{k_{max}} \omega_k, $
since matter is practically transparent for high-frequency modes.
Note some peculiarities of the volume Casimir effect for solid ball:
\begin{itemlist}
\item
Unlike the surface Casimir effect, it
has a little sensitivity to geometry of the boundary surface and its total
energy is additively connected with the volume, occupied by material.
\item
It gives negative contribution and depends upon the fourth power of
interatomic separation $d$, while the usual matter is proportional $d^{-3}$.
Hence, when compressing the matter, it increases faster than matter density.
\end{itemlist}

There exist a bound when the matter density $E_m=md^{-3}$ (m is the mass of
particle) and the vacuum contribution $\Delta E_{vac}$ are equal and can be
compensated:
\begin{equation}
E_{tot}=E_m +\Delta E_{vac} = md^{-3} - \pi^2 d^{-4}/8 \approx 0.
\label{comp}
\end{equation}
This bound corresponds to `superclosely packed' particles when the
separations between them are $d\approx\pi ^2/8m$, of the order of their
Compton length. Further compression of the matter could be impossible
without destruction of particles according to the principles of QED.

It shows that the volume Casimir effect can be strong for island
systems of superdense matter, and, in the absence of the
other withstanding effects, the existence of a superdense pseudovacuum
state of matter corresponding to $E_{tot}\approx 0$ could be expected.

\subsection{One calculable and experimentally verified model}

The considered above model is based on the rough estimations. However,
it can be modified turning into a finite, calculable and experimentally
verified model.
 The Casimir effect with parallel plates is computable, it does not
contain the connected with structure of
matter cut-off parameter and is experimentally supported.  The negative
zero-point energy contribution to intermediate space between plates is
\begin{equation} \Delta E_{vac}= - \pi ^2 /720 a^{4} \label{bat}
\end{equation}
where $ a$ is interplate separation.

Solid ball can be modelled by a battery of parallel plates.
In this case the volume Casimir effect depends on separations between
plates $a$ and can  be experimentally checked.

On the other hand, in the assumption that the thickness of the plates is
much
smaller than interplate separations, the total vacuum contribution will be
determined by the value (\ref{bat}).
In fact, form of the model (ball or maybe box) is not essential
since the dependence on the form can be attributed to the corresponding
surface Casimir effect.

\subsection{Higgs field model for superconducting ball}

Instead of the step cut-off function for $k>k_{max}$, a smooth decreasing of
the permitivity upon the frequency can be considered \cite{Mil,SDM}. For
high
frequencies $\epsilon (\omega) = 1 - 4\pi N e^2/m\omega^2,$ where $e$ is
charge of
particles providing conductivity, $m$ their mass, $N$ the density.
Maxwell equations yield then for the transverse components of vector
potential
\begin{equation}
( \Box + M^2 (r) ) U =0,
\label {1}
\end{equation}
where
\begin{equation}
M^2(r)= 4\pi N e^2/m
\label {M}
\end{equation}
inside the conducting ball, and $M=0$ in external region.
If we set $\phi(r) = M(r)/e$, the equation (\ref{1}) can be treated as
consequence of the Higgs model of superconductivity where electromagnetic
field acquires mass $M$, interacting with the Higgs field $\Phi(r)\ne 0$
inside the ball.
The low-frequency modes with $\omega<M$ are `pushed out' of the ball, and
the solutions inside the ball fall off exponentially.
The effective energy-momentum tensor depends upon coordinate
$r$ and is determined by the regularization relative to the region outside
the ball
$T^{\mu\nu}_{eff}(r)= <0\vert T^{\mu\nu}(r)\vert 0> -
<0\vert T^{\mu\nu}(\infty)\vert 0>.$
The substraction has to be carried out `mode in mode' \cite{DW}, when for
separate mode we have
$T_{\mu\nu}(U_m,U_m^\ast)= \partial _\mu U_m \partial _\nu U_m^\ast
-\frac 12 g_{\mu\nu} \partial _\rho U_m \partial ^\rho U_m^\ast
-\frac 12 M^2(r)g_{\mu\nu} \vert U_m \vert ^2.$
The orthonormal set of basis functions $U_m(x)$ is determined by the smooth
matching of the solutions outside and inside the ball.
As a result \cite{Bur} the regularized value of energy density inside the
ball integrated over $\omega$ to $\omega_{max}$ is
\begin{equation}
\Delta E_{vac} (\omega_{max}) =
- M^4/32 \pi^2  - M^2(\omega ^2_{max} -M^2)/8\pi^2.
\label {Emax}
\end{equation}
It is quadratically divergent. This means that matter influences not only
the low-frequency modes with $\omega<M$ but also the high-frequency modes
penetrating into the matter, thus leading to their energy density decrease.
However, simultaneously the increasing of material energy takes
place in the process of field-particles interaction \cite{BLP}. For
plane waves with momentum $k$ and polarizations
$\alpha$ the one-particle energy change is
$\delta \epsilon =\sum_{k \alpha} \frac {e^2} {2m\omega}
\bar{\vert A_{k\alpha}^2\vert}.$
The resulting alteration of material energy density will also be divergent
\begin{equation}
\Delta E_{int} =\frac {2\pi e^2 N} {m(2\pi)^3} \int_M^{\omega_{max}}
\frac {dk^{(3)}}{\omega}= M^2(\omega ^2_{max} -M^2)/8\pi^2
\label {Eint}
\end{equation}
and compensates the divergent term of (\ref{Emax}) leading to the finite
result
\begin{equation}
\Delta E_{vac} = - M^4/32 \pi^2 = - \frac 12 N^2 e^4/m^2.
\label {EHvac}
\end{equation}
Cancelation of divergent terms is consequence of a hidden supersymmetry in
this model.

\section{Supersymmetric Superconducting Ball Model}

In previous model the Higgs field $\phi(r)$ was introduced by hand, filling
the ball interior.
In the similar soliton-like solutions, ``lumps" and ``Q-balls",
the scalar field $\phi(r)$ concentrates inside the island systems that is
controlled by a special nonlinear potential $V(\phi)$.
The systems of this kind we will call  "type A" -systems.
In the Higgs model the scalar field $\phi _A (r)$ is complex and
gives mass $M(r) = e^{-1}\phi_A (r)$ to electromagnetic field
$A_\mu \quad (F_{A\mu\nu})$ inside the ball
\begin{equation}
L_A=-\frac 14 F_A^{\mu \nu }F_{A\mu \nu} -
2(D^\mu \phi )\overline {( D_\mu  \phi  )} - V{\phi},
\label {Higgs}
\end{equation}
where $D_A^\mu = \nabla_\mu +ie A_{\mu}.$
The known ball-like models do not contain gauge fields, and
it is apparently not possible to construct the consistent type A
model with long range external electromagnetic field and finite total
energy.
\par
On the other hand, the known models based on
the Higgs field (like  bags, domain walls, bubbles and strings)
display usually a ``dual"  behavior (say ``type B") in the sense that the
field $\phi_B (r)$ forms a bag or stringlike cavity  going to a constant
vacuum value $\phi _B = \phi _{vac} \ne 0$ in external region.
The  finite energy demand is usually satisfied, however the gauge
field $B_\mu$ acquires mass and becomes short range in external region.
We call this system as``type B" one.  It corresponds to $A\to B$
substitution in (\ref{Higgs}).
Thus, none of the systems $A$ or $B$ can separately provide
necessary demands for description of superconducting ball.
\par
The exit can be found out in the Witten $U(I) \times \tilde U(I) $ field
model which was used for description the cosmic superconducting strings
\cite{Wit}. We use the supersymmetric version of the
Witten field model \cite{Mor} which yields PBS-saturated solutions
\cite{Bag} and
supersymmetric vacuum states which are free from quantum corrections.
\par
The Witten field model represents a doubling of the Higgs model and
contains the both sectors: type A and B.
The long range electromagnetic fields $A_\mu$
acquires mass from field $\phi\equiv\phi_A$ filling the ball, and the
gauge field $B_{\mu}$ is confined inside the bag formed by scalar field
$\sigma\equiv\phi _B $.
 Supersymmetric version of the Witten field
model has effective Lagrangian
$L=-2(D_A^\mu \phi  )\overline {( D_{A\mu}  \phi  )}
-2(D_B^\mu \sigma )
(\overline {D_{B\mu} \sigma} )-\partial ^\mu Z \partial _\mu
\bar Z \nonumber
-\frac 14F_A^{\mu \nu }F_{A\mu \nu }-
\frac 14 F_B ^{\mu \nu }F_{B\mu \nu}-V (\sigma, \phi, Z),$
where the potential
$V= \sum _{i=1} ^5 \vert \partial_i W \vert ^2$.
The superpotential $W(\Phi ^i)$ is a holomorphic function of
the fife complex chiral fields
$\Phi ^i =\{ Z,\phi, \bar\phi, \sigma, \bar \sigma \}$,
\begin{equation}
W=\lambda {Z}(\sigma \bar\sigma -\eta ^2) + ( c Z+m ) \phi\bar \phi.
\label{bur-SW}
\end{equation}
\par
In the effective Lagrangian the "bar" is identified with complex
conjugation, so there are really only three independent scalar fields and
the "new" ( neutral ) fields Z provides the synchronization of the phase
transition.
The supersymmetric vacuum states corresponding to the lowest value of the
potential are determined by the conditions $\partial_i W=0$.
It yields two supersymmetric vacuum states with $V=0$:
\begin{equation}
I ) \qquad Z=0;\quad \phi=0 ;\quad \vert\sigma\vert=\eta ;\quad W=0,
\label{bur-true}
\end{equation}
we set it for external vacuum; and
\begin{equation}
II ) \qquad Z=-m/c;\quad \sigma=0; \quad \vert \phi \vert =\eta
\sqrt{\lambda/c};\quad W=\lambda m\eta ^2/c= m\phi ^2,
\label{bur-false}
\end{equation}
we set it as a state inside the bag.
\par
The treatment of the gauge field $A_\mu$ and $B_\mu$ in $B$ is similar
in many respects because of the symmetry between  $A$ and $B$ sectors
allowing one to consider the state I ) in outer region as
superconducting one in respect to the gauge field $B_\mu$, which provides
confinement of the $B_{\mu}$ field inside the bag. \footnote{There
are some evidences that in B-sector a ``dual" type of superconductivity has
to be realized.}
\par
One can check the phase transition in the planar wall approximation
neglecting the gauge fields.
Using the Bogomol'nyi
transformation one can represent the energy density as follows
\begin{equation}
\rho =
\frac{1}{2} \delta _{ij}\lbrack  \Phi ^i,_z + \frac {\partial W}{\partial
\Phi ^j}\rbrack
\lbrack  \Phi ^j,_z + \frac {\partial W}{\partial\Phi ^i}\rbrack
- \frac{\partial W}{\partial \Phi ^i}\Phi ^i,_z,
\label{trbog}\end{equation}
where the last term is full derivative.
Then, integrating over the wall depth $z$ one obtains for the surface
energy density of the wall
$\epsilon=\int _0 ^\infty \rho dz =\frac 12 \int\Sigma _i
( \Phi ^i,_z + \frac {\partial W}{\partial\Phi ^i})^2 dz + W(0) -W(\infty).$
The minimum of energy is achieved when the first-order Bogomol'nyi
equations $\Phi ^i,_z + \frac {\partial W}{\partial\Phi ^i} =0$ are
satisfied.
Its value is given by
$\epsilon =W(0)-W(\infty)= m \phi _{A(in)}^2 $.
Therefore, this domain wall is BPS-saturated solution, and
corresponding vacuum states are supersymmetric and do not acquire quantum
corrections. Supersymmetric vacuum state inside the ball has energy density
$\rho =V/2=0$, in spite of the nonzero density of scalar field there,
displaying analogue with (\ref{comp}).
In supergravity this problem is connected with the regular black hole and
particlelike models \cite{Bag,SBH} when singularity is replaced by  a
superconducting source. The extra contribution to potential in $N=1$
supergravity can yield negative vacuum energy \cite{CvSol}
\begin{equation}
\rho =- \frac 3 2  W \bar W =-\frac 3 2  m^2 \phi  ^4 ,
\label{sugra}
\end{equation}
(here $m$ is parameter of supersymmetric model) leading to AdS vacuum
state inside the source.
\par
\section{Conclusion}
The negative contribution of the Casimir zero-point energy for
(super)conducting ball turns out to be finite and calculable in represented
models, showing its importance for island systems with superdense matter.
\par
\nonumsection{Acknowledgments}
Author would like to thank Organizing Committee of the Conference, and in
particular Prof. Michael Bordag, for very kind invitation and financial
support.

\eject
\end{document}